
\input harvmac
\overfullrule=0pt
\def\nl{\hfill\break}
\def\bn{{\bf n}}
\def\mod{~{\rm mod}~}

\def\ch{{\rm ch}}

\def\cS{{\cal S}}
\def\cO{{\cal O}}
\def\bn{{\bf n}}
\def\mod{~{\rm mod}~}

\def\am{^{(a)}_m}
\def\Log{{\rm Log}}
\def\ti{\tilde}

\Title{HUTP-92/A069}
{\vbox{\centerline{Characters in Conformal Field Theories}\nl
       \centerline{from Thermodynamic Bethe Ansatz}}}
\centerline{Atsuo Kuniba$^{1,}$
\footnote{$^a$}
{e-mail: kuniba@math.sci.kyushu-u.ac.jp}
, Tomoki Nakanishi$^{2,}$
\footnote{$^b$}
{e-mail: nakanisi@string.harvard.edu}
\footnote{\null}{Permanent Address:
Department of Mathematics, Nagoya University,
Nagoya 464 Japan}, Junji
Suzuki$^{3,}$
\footnote{$^c$}
{e-mail: jsuzuki@tansei.cc.u-tokyo.ac.jp}}
\vskip0.5cm
\bigskip\centerline{$^1$Department of Mathematics, Kyushu University}
\centerline{Fukuoka 812 JAPAN}
\bigskip\centerline{$^2$Lyman Laboratory of Physics, Harvard University}
\centerline{Cambridge, MA 02138 USA}
\bigskip\centerline{$^3$Institute of Physics, University of Tokyo, Komaba}
\centerline{Meguro-ku, Tokyo 153 JAPAN}

\vskip .3in
Abstract. We propose a new $q$-series formula for a character of
parafermion conformal field theories associated to
arbitrary non-twisted affine Lie algebra $\widehat{g}$.
We show its natural origin from a thermodynamic Bethe
ansatz analysis including chemical potentials.

\Date{12/92}

%
\nref\Ki{A.N.Kirillov, Zap.Nauch.Semin.LOMI {\bf 164} (1987) 121 and private
communications}
\nref\BRIJ{V.V.Bazhanov and Yu.N.Reshetikhin, Int.J.Mod.Phys. {\bf A4}
(1989) 115}%
\nref\BRJP{V.V.Bazhanov and Yu.N.Reshetikhin, J.Phys. {\bf A23} (1990) 1477}
\nref\Ku{A.Kuniba, Nucl.Phys. {\bf B389} (1993) 209}
\nref\ABF{G.E.Andrews, R.J.Baxter and P.J.Forrester,
J.Stat.Phys. {\bf 35} (1984) 193}
\nref\DJKMONP{E.Date, M.Jimbo, A.Kuniba, T.Miwa and M.Okado,
Nucl.Phys. {\bf B290} [FS20] (1987) 231; Adv.Stud.in Pure Math.
 {\bf 16}
(1988) 17}
\nref\JKMO{M.Jimbo, A.Kuniba, T.Miwa and M.Okado,
Commun. Math. Phys.{\bf 119} (1988) 543}
\nref\FZparaf{V.A.Fateev and A.B.Zamolodchikov,
Sov.Phys.JETP {\bf 62} (1985) 215}
\nref\Ge{D.Gepner, Nucl.Phys. {\bf B290} [FS20] (1987) 10}
\nref\KN{A.Kuniba and T.Nakanishi, Mod.Phys.Lett. {\bf A7}
(1992) 3487}
\nref\BPZ{A.A.Belavin, A.M.Polyakov and A.B.Zamolodchikov,
	Nucl.Phys. {\bf B241} (1984) 333.}
\nref\YY{C.N.Yang and C.P.Yang, J.Math.Phys. {\bf 10} (1969) 1115}
\nref\Albzamo{Al.B.Zamolodchikov, Nucl.Phys. {\bf B342} (1990) 695,
{\bf B358} (1991) 497,  Phys.Lett. {\bf B253} (1991) 391}
\nref\KM{T.R.Klassen and E.Melzer, Nucl.Phys. {\bf B338} (1990) 485;
{\bf B350} (1991) 635}
\nref\KP{V.G.Kac and D.H.Peterson, Adv. Math. {\bf 53} (1984) 125}
\nref\RS{B.Richmond and B.Szekeres, J.Austral.Math.Soc. Series A
 {\bf 31} (1982) 362}
\nref\NRT{W.Nahm, A.Rechnagel and M.Terhoeven, ``Dilogarithm Identities in
Conformal Field Theory", Bonn preprint (1992)}
\nref\Ter{M.Terhoeven, ``Lift of Dilogarithm to Partition
Identities", Bonn preprint (1992)}
\nref\LP{J.Lepowski and M.Primc, Contemporary Math. 46 AMS, Providence (1985)}
\nref\Kac{V.G.Kac, {\sl Infinite dimensional Lie algebras},
(Cambridge University Press, 1990)}
\nref\KW{V.G.Kac and M.Wakimoto, Adv. Math. {\bf 70} (1988) 156}
\nref\KKMM{R.Kedem, T.R.Klassen, B.M.McCoy and E.Melzer,``
Fermionic Quasi-Particle Representations for Character of
$(G^{(1)}_1\times G^{(1)}_1)/ G^{(1)}_2$", ITP preprint (1992)}
\nref\And{G.E.Andrews, {\sl The Theory of Partitions},
Addison-Wesley ,1976}
\nref\DJKMO{E.Date, M.Jimbo, A.Kuniba, T.Miwa and M.Okado,
Lett.Math.Phys. {\bf 17} (1989) 69; Adv.Stud.in Pure Math. {\bf 19}
(1989) 149}
\nref\FNO{B.L.Feigin, T.Nakanishi and H.Ooguri, Int.J.Mod.Phys. {\bf A7},
Suppl. 1A (1992) 217}
\nref\Fen{P.Fendley, Nucl.Phys. {\bf B374} (1992) 667}
\nref\DVK{H.J.de Vega and M.Karowski, Nucl.Phys. {\bf B285}[FS19] (1987)
619}
\nref\FZPLB{V.A.Fateev and Al.B.Zamolodchikov, Phys.Lett. {\bf B271} (1991)
91}
\nref\Maparaf{M.J.Martins, Phys.Rev.Lett. {\bf 65} (1990) 2091}
\nref\Rav{F.Ravanini, Phys.Lett. {\bf B282} (1992) 73}
\nref\Abzamo{A.B.Zamolodchikov, Int.J.Mod.Phys. {\bf A4} (1989) 4235}
\nref\BCN{H.W.J.Bl\"ote, J.L.Cardy
and M.P.Nightingale,  Phys.Rev.Lett. {\bf 56} (1986) 742}
\nref\Aff{I.Affleck, Phys.Rev.Lett. {\bf 56} (1986) 746}
\nref\Mae{M.J.Martins, Phys.Rev.Lett. {\bf 67} (1991) 419}
\nref\KLP{A.Kl\"umper and P.A.Pearce, J.Stat.Phys. {\bf 64}
(1991) 13; Physica {\bf 183A} (1992) 304}
\nref\KMe{T.R.Klassen and E.Melzer, Nucl.Phys. {\bf B370} (1992) 511}
\nref\KNS{A.Kuniba, T.Nakanishi and J.Suzuki, in preparation}

\beginsection 1. Introduction

Recently new aspects in conformal field theories (CFTs)
are being recognized through
studies of thermodynamic limit of integrable models
such as 1d quantum spin chains and $(1+1)$d factorized
scattering systems.
In these analysis, the Rogers dilogarithm function
\eqn\dilog{
L(x)=-{1 \over 2} \int_{0}^{x} \left(
        {\log (1-y) \over y}+{\log y \over 1-y}
        \right) dy
}
plays a key role that connects thermodynamic quantities
in those models to the CFT data, most notably, central
charges and scaling dimensions.
For example, the following
conjecture emerged \Ki,\BRIJ,\BRJP,\Ku\ from the
restricted solid-on-solid (RSOS) type \ABF,\DJKMONP,\JKMO\
spin chains:
\eqn\ccc{
{\ell \dim g \over g^\vee + \ell} -r=
{6\over \pi^2}\sum_{(a,m)\in G} L(f^{(a)}_m),
}
where the lhs is the central charge $c_{\rm PF}$
of the parafermion (PF) CFT \FZparaf,\Ge\
associated to an affine Lie algebra
$\widehat{g}$ with
rank $r$, level $\ell$ and dual Coxeter number $g^\vee$.
(See \KN\ for a generalization of \ccc\ including the scaling dimensions.)
The set $G$ is given by (5) and
$f^{(a)}_m$ is the unique solution to the simultaneous algebraic
equation in the range
$0<f^{(a)}_m <1$,
\eqn\feq{
f^{(a)}_m=\prod_{(b,k) \in G} (1-f^{(b)}_k)^{
K^{mk}_{ab}}\quad {\rm for }\,\, (a,m) \in G,
}
\eqn\kdef{
K^{mk}_{ab}=
\left( \min(t_b m ,t_a k  )
-{mk \over  \ell} \right)
(\alpha_a | \alpha_b),
}
with the notations specified later.
Needless to say, the equation of such form as well as the
appearance of the dilogarithm are reflecting rich
structures encoded
in the integrable models.
Eq.\ccc~is thereby connecting the two fundamental ingredients;
the CFT data \BPZ\ on the lhs which is of affine Lie algebraic origin
and
the intricate formula on the rhs occurring from thermodynamics of the
integrable models.
\par
The purpose of this Letter is to put forward such a
connection even further based on the thermodynamic
Bethe ansatz (TBA) \YY,\BRIJ,\BRJP,\Ku,\Albzamo,\KM.
We shall propose a new $q-$series formula for a PF character,
which is essentially equivalent to
a string function \KP\ of any non-twisted affine
Lie algebra $\widehat{g}$ at any level $\ell \in {\bf Z}_{\ge 1}$.
It has a surprisingly simple form and seems to reveal
an interesting
structure of the PF modules.
When $q \rightarrow 1^-$, the $q-$series formula leads to \ccc\
by comparing the asymptotics on both sides with the method of
\RS.
Thus our new proposal (9) may be viewed as a
``lift" of \ccc\ to a PF character formula in the sense of
\NRT,\Ter.
More importantly, we point out that the $q-$series
formula arises naturally from the spectra of the
TBA-originated effective central charge \KN\ involving dilogarithms.
The key is to observe a one to one correspondence
between the independent  states in the Hilbert space
of the PF CFT
and the ways of analytic continuations of the
dilogarithm.
The idea provides a new insight toward
a structural correspondence between CFTs and TBA hence
its presentation also consists of our main aim
in this Letter.
We remark that for the special case
$\widehat{g} = A^{(1)}_1$, our $q-$series formula
coincides with that in \LP.
\vskip5mm \noindent
{\bf 2. New $q-$series formula }
\par\noindent
Let $g$ denote one of the classical simple Lie algebras
$A_r (r \ge 1), B_r (r \ge 2), C_r (r \ge 1),
D_r (r \ge 4), E_{6,7,8}, F_4$ and $G_2$.
We write $r = {\rm rank }\, g$ and $\widehat{g}$ to mean
the non-twisted affinization of $g$ \Kac.
Let $\Delta$, $\Delta_+$, $\Pi$, $h$, $(\cdot | \cdot)$
denote the root system, the set of positive roots,
the set of the simple
roots,
the Cartan subalgebra,
the invariant form on $g$,
respectively.
The spaces $h$ and $h^*$ are identified
via the form $(\cdot | \cdot)$.
We employ the normalization
$\vert$long root$\vert^2=2$ and set
$t_a=2/(\alpha_a|\alpha_a), \,
\alpha_a^{\vee} = t_a \alpha_a$ for each simple root
$\alpha_a$, where the nodes $1 \le a \le r$
on the Dynkin diagram are enumerated according to \Kac.
The root lattice
$Q=\bigoplus {\bf Z}\alpha_a$, the coroot lattice
$Q^{\vee}=\bigoplus {\bf Z}\alpha_a^{\vee}$ and
the weight lattice $P=(Q^\vee)^*$ are as usual.
We find it convenient to label the weights
of $\widehat{g}$ (mod null root) by its projection
onto the classical part $P$.
Throughout the Letter we
fix an integer $\ell \in {\bf Z}_{\ge 1}$ and
put $\ell_a = t_a \ell$ and
\eqn\gdef{
G=\{ (a,m)| 1 \leq a\leq r, 1\leq m \leq \ell_a-1,\, a,m \in {\bf Z}\}
}
following \Ku, \KN.
\par
Let $L^\Lambda$ denote the integrable
$\widehat{g}$-module
having a level $\ell$ dominant integral weight
$\Lambda$ as the highest weight \Kac.
One can fit the action of the (homogeneous) Heisenberg
algebra $\widehat{a}$ of rank $r$ on  $L^{\Lambda}$ \LP,\Ge.
The algebra $\widehat{a}$ has a basis $\{ a^x_{n} |
x \in \Pi, n \in {\bf Z} \}\cup\{id\}$.
The irreducible module $\Omega^\Lambda$
of PF algebra is isomorphic to
the subspace of $L^\Lambda$ consisting of
the vectors $v$ such that\eqn\anh{
       a^x_n v=0~~~~{\rm for}~x \in \Pi,~~~n>0.
}
The space admits the weight space decomposition
\eqn\wsd{
        \Omega^\Lambda = \bigoplus_{{
\lambda \in P \atop \lambda \equiv \Lambda ~ {\rm mod}~
Q }} \Omega^\Lambda_\lambda.
}
The PF currents $\psi^\alpha_n$ ($\alpha \in \Delta$),
which commute with the operators
$a^x_{\pm n} \, (n \in {\bf Z}_{\ge 1})$,
map the elements in $\Omega^\Lambda_\lambda$ into another sector
$\Omega^\Lambda_{\lambda+\alpha}$.
The character
of $\lambda$-weight sector $\Omega^\Lambda_\lambda$
(with variable $q$) is given by \KP,\Ge
\eqn\char{
\ch (\Omega^\Lambda_\lambda)=\eta(q)^r c^\Lambda_\lambda(q),
}
where $c^\Lambda_\lambda(q)$ is a string function of $\widehat{g}$
at level $\ell$  and
$\eta(q)$ is the Dedekind eta function.
The string function is by
definition the character of the (graded)
$\lambda$-weight subspace of
$L^\Lambda$, which is of fundamental importance.
So far its explicit formula is not known for general
$\widehat{g}$ and $\ell$ although several expressions are
available in some cases \KP,\KW,\LP.
Let $\bar\Omega^{\Lambda}$  be the quotient of
the space $\Omega^{\Lambda}$ by the identification
$\Omega^{\Lambda}_\lambda \sim
\Omega^{\Lambda}_{\lambda+\ell Q^{\vee}}$, and
the Hilbert space of the chiral half of the PF CFT
corresponds to
the direct sum of $\bar\Omega^{\Lambda}$'s.
\par
{}From now on we shall exclusively consider
{\it the vacuum module} $\Omega^{0}$ case and
propose the following character formula
for each $\lambda$-sector ($\lambda \in Q$):
\eqn\propa{
\ch (\Omega^{0}_\lambda)=q^{-c_{{\rm PF}}/24}
\sum_{\lambda(\bn)\equiv
\lambda\mod \ell Q^{\vee}} {q^{{\cal K}({\bf n})}
\over (q)_{\bn}},
}
\eqn\propb{
{\cal K}({\bf n}) = {1 \over 2}
\sum_{\scriptstyle (a,m) \in G \atop \scriptstyle (b,k) \in G}
K^{m\, k}_{a\, b} n^{(a)}_m n^{(b)}_k,
}
\eqn\propc{
(q)_{\bn} = \prod_{(a,m) \in G}(q)_{n^{(a)}_m} , \quad
(q)_k = \prod_{j=1}^k (1-q^j).
}
Here the summation in \propa\
 runs over
the vectors \eqn\nvec{
\bn =( n^{(a)}_m )_{(a,m) \in G} \in ({\bf Z}_{\geq 0})^{|G|}
}
under the indicated
restriction $\lambda(\bn )\equiv \lambda \mod \ell Q^\vee$ with
\eqn\lamn{
 \lambda(\bn)=\sum_{(a,m)\in G} m n^{(a)}_m
\alpha_a,
}
which is compatible with the
invariance property $c^{\Lambda}_{\lambda}=
c^{\Lambda}_{\lambda +\ell Q^{\vee}}$.
Under the above restriction, it can be easily shown that
the rhs of \propa\
 contains only non-negative integer powers of
$q$ up to an overall factor $q^p$ with
$p \equiv -{c_{\rm PF} \over 24}-{\vert \lambda \vert^2 \over 2\ell}$
mod ${\bf Z}$.
The character of the space $\bar\Omega^{0}$
is now given by the same formula \propa\
but {\it without} any restriction on the $\bn-$sum other than \nvec.
At present, a proof is not known for \propa\
 for general
$\widehat{g}$ and $\ell$.
However one can verify several cases directly and observe a
wealth of consistency as we shall see below.
For $\widehat{g} = A^{(1)}_1$, some generalizations into
different directions have also been conjectured
in \Ter,\KKMM.
\par
Firstly, \propa\ is indeed valid for
$(\widehat{g},\ell) = (A^{(1)}_1,{\rm general})$
as it coincides \Ter\ with the formula in \LP.
So is the case $(\widehat{g},\ell) = (B^{(1)}_r,1)$ with $r$
general, where one can actually compute the $\bn-$sum by means of
eq.(2.2.6) in \And\ and compare the result with that in \KP.
The case $(\widehat{g},\ell) = (G^{(1)}_2,1)$ can also be proved
since the $q-$series \propa\ then
reduces to that for $(A^{(1)}_1,3)$ (cf.\KP).
Not to mention, \propa\
 is trivially ture for
$\widehat{g} = A^{(1)}_r, D^{(1)}_r$ and $E^{(1)}_{6,7,8}$
with $\ell = 1$ when the PF module becomes 1 dimensional.
Secondly, we have generated the low power terms in
\propa\
 by computer and checked agreements
with the known results on the string functions for
$(\widehat{g},\ell) = (A^{(1)}_{2,3},2,3)$ \DJKMO,
$(C^{(1)}_{3,4},1),(F^{(1)}_4,1)$ and $(E^{(1)}_8,2)$
\KP,\KW.
For instance in the last example, the rhs of \propa\
with $\lambda = \Lambda_1$ yields
$q^{3/16}(1 + 29q + 288q^2 + 1878q^3 + \cdots)$.
This agrees with the $E^{(1)}_8$ level 2 result
$b^{2\Lambda_0}_{\Lambda_1}$
given by eq.(4.4.3a) and Proposition 4.4.1(e) in \KW\
as an order 9 polynomial of the Virasoro characters.
($\Lambda_i$ denotes the $i-$th fundamental weight.)
One may substitute \char\ and \propa\
 into the character formula \KP
\eqn\kmch{
\ch(L^0)=
\sum_{\lambda \in  Q/\ell Q^{\vee}}
 c^0_{\lambda}(q) \Theta_{\lambda,\ell}(z,q),
\quad
\Theta_{\lambda,\ell}(z,q)=\sum_{\gamma \in \lambda +\ell Q^\vee}
 q^{|\gamma|^2/2\ell} e^{-2\pi i (\gamma \vert z)}
}
under the principal specialization of $z$.
We have then checked that the resulting $q-$series for $\ch(L^0)$
indeed fulfills the known
factorization property \KP\ up to some power
for many examples including
$(\widehat{g},\ell) = (A^{(1)}_{2,3},3), (B^{(1)}_3,2,3),
(D^{(1)}_4,2,3),$ $
(F^{(1)}_4,2)$ and $(G^{(1)}_2,2)$.
Thirdly,
if \propa\
 is true, then
\eqn\powg{
{\rm min}\{ {\cal K}(\bn) \mid \nvec \, {\rm and }\,
\lambda(\bn) \equiv \lambda \, {\rm mod }\,\ell Q^{\vee}\}
= n^{0}_\lambda -{ |\lambda|^2 \over 2 \ell}
}
must hold by comparing the leading powers on
both sides.
Here $n^0_\lambda$ is the minimum eigenvalue of the Virasoro
operator $L_0$ in the $\lambda$-weight subspace of $L^0$ and
is equal to the minimum number of roots to express $\lambda$
as their sum if it is possible within
$\ell$ roots.
Our quadratic form ${\cal K}(\bn)$ \propb\
 has the consistent
property to it since
\eqn\powt{
{\rm min}\{ {\cal K}(\bn) \mid \nvec \, {\rm and }\,
\lambda(\bn) \equiv \lambda\, {\rm mod }\, \ell Q^{\vee} \}
= 1 -{ |\lambda|^2 \over 2 \ell}
}
is valid for any positive root $\lambda$, which is a special case of
\powg.
Finally, we remark that \propa\
 is also consistent
in that it leads to the dilogarithm conjecture \ccc\
 by
comparing the asymptotics on both sides as $q \rightarrow 1^-$.
To see this,
we firstly note that the leading divergence of the lhs in \propa\
 is
$({\bar q})^{-c_{\rm PF}/24} \, ({\bar q} = e^{-2\pi i/\tau})$ when
$q = e^{2\pi i \tau} \rightarrow 1^-$ \KP.
As for the rhs, one can apply the
argument in \RS,\NRT\ to get a crude estimate
$({\bar q})^{-\sum L(f^{(a)}_m)/4\pi^2}$, from which
\ccc\
 follows.
In particular, \feq\
 arises
essentially from the ``saddle-point condition"
$q^{n^{(a)}_m} = 1 - f^{(a)}_m$ with respect to $n^{(a)}_m$.
\par
Before closing this section, let us
discuss how our $q$-series \propa\
 will
indicate a basis structure in the
PF module in the light of the earlier works \LP,\FNO.
The space $\Omega^{0}$ is certainly spanned by the vectors
\eqn\span{
T^{\gamma} \psi^{\beta_1}_{-k_1} \cdots \psi^{\beta_j}_{-k_j} v_0
\quad (\gamma \in \ell Q^{\vee},
\beta_i \in \Delta_+, k_i \in {\bf Z}_{\ge 1}), }
where $v_0$ is the highest weight vector and
$T^{\gamma}$ is the translation isomorphism $T^{\gamma}:
\Omega^{0}_{\lambda} \rightarrow
 \Omega^{0}_{\lambda+\gamma}$.
Furthermore
by introducing a lexicographic
ordering in
this set, we can choose Poincar\'e-Birkhoff-Witt
type vectors among them as a spanning set of $\Omega^0$.
To illustrate the idea let us take the example
$(\widehat{g},\ell) = (A^{(1)}_2,2)$ with
$\lambda = \alpha_1 + \alpha_2 \in \Delta$ and consider the
$n^{(1)}_1=n^{(2)}_1=1$ term ${q^{1/2} \over (q)_1(q)_1}$
in \propa\
(apart from $q^{-c_{\rm PF}/24}$).
In view of \lamn\
 and the restriction
$\lambda(\bn )\equiv \lambda \mod \ell Q^\vee$,
it corresponds to the character of the
subspace of $\Omega^{0}_{\alpha_1 + \alpha_2}$
spanned by the vectors
\eqn\psiv{
\matrix{
\hfill  \psi^{\alpha_1 + \alpha_2}_{-k}v_0&& (k\geq 1),\hfill \cr
\hfill  \psi^{\alpha_{1}}_{-k_1}
\psi^{\alpha_{2}}_{-k_2}v_0 && (k_1> k_2 \geq 1), \hfill\cr
\hfill  \psi^{\alpha_{2}}_{-k_1}
\psi^{\alpha_{1}}_{-k_2}v_0 && (k_1 \geq k_2 \geq 1), \hfill\cr
}
}
since their contributions amount to it as
\eqn\cont{
q^{-1/2}\left(
{q \over (q)_1}+{q^3 \over (q)_2}+{q^2 \over (q)_2}\right)
={q^{1/2} \over (q)_1 (q)_1}.
}
Here the prefactor $q^{-1/2}$ comes from
$-|\alpha_1 + \alpha_2|^2/2\ell
=-1/2$.
In general non-trivial relations exist among the operators
$\prod_i \psi^{\beta_i}_{-k_i}$
if $(\sum \beta_i|\Lambda_j) \geq \ell$ for some fundamental weight
$\Lambda_j$, hence one must eliminate some spanning vectors
to get a real basis.
We leave it as an interesting future problem.
\vskip5mm \noindent
{\bf 3. Origin from TBA}
\par\noindent
%
Our proposal \propa\
 for the PF character has stemmed from
an analysis based on the TBA type integral equation in \KN
\eqn\tba{
R {\cal M}_a\hbox{ch}u =  \pi i D^{(a)}_m + \epsilon^{(a)}_m(u) +
\sum_{(b,k) \in G} \int_{- \infty}^\infty
dv \, \Psi^{m k}_{a b}(u - v)\, \hbox{log}
\bigl(1 + \hbox{exp}(-\epsilon^{(b)}_k(v)) \bigr),
}
which represents interacting ``pseudo particles" with
energy $\epsilon^{(a)}_m(u)$ labeled by $(a,m) \in G$.
Here, ${\cal M}_a > 0$, $\pi i D^{(a)}_m$ and $R$ are independent
of the rapidity $u$ and stand for the mass,
the chemical potential (cf.\ \Fen) and
the system size corresponding to the inverse temperature in TBA.
The integration kernel $\Psi^{m k}_{a b}(u)$
decays rapidly when $\vert u \vert \rightarrow \infty$ and
has been specified in eq.(18) of \KN.
Here we will not need its explicit form but the properties
\eqn\ker{
\int_{-\infty}^{\infty} \Psi^{m k}_{a b}(u) du
= \delta_{a b} \delta_{m k} - K^{m k}_{a b},\quad
\Psi^{m k}_{a b}(u) = \Psi^{m k}_{a b}(-u) = \Psi^{k m}_{b a}(u).
}
Eq.\tba\
 has a similar form to many earlier examples of the
TBA equations \DVK,\Albzamo,\FZPLB,\KM,\Maparaf,\Rav\ and
is a candidate describing
a massive  deformation of the level $\ell$ $\widehat{g}$ PF CFT by a
certain relevant operator \Abzamo.
Actually, one can apply the standard TBA technique
to show that the free energy
\eqn\free{
F(R) = -{1 \over 2\pi}
\sum_{a=1}^r {\cal M}_a \sum_{m=1}^{\ell_a - 1}
\int_{- \infty}^\infty du \, \hbox{ch}u \, \hbox{log}
\bigl(1 + \hbox{exp}(-\epsilon^{(a)}_m(u)) \bigr)
}
has the ultraviolet (UV) asymptotics
\eqn\fbeh{
F(R) \simeq -{\pi c \over 6 R} \,\, \hbox{ as } R \rightarrow 0,
}
\eqn\cfun{
{\pi^2 \over 6}c = \sum_{(a,m) \in G}
\bigl(L(f^{(a)}_m) - {\pi i\over 2}D^{(a)}_m \log(1-f^{(a)}_m)
\bigr),}
\eqn\dter{
\pi i D^{(a)}_m = \log f^{(a)}_m -
\sum_{(b,k) \in G} K^{m k}_{a b} \log (1-f^{(b)}_k).}
Eq.\fbeh\ is a characteristic behavior of the CFT
with the central charge $c$ \BCN,\Aff.
In the derivation, we have used \ker\ and put
$\epsilon^{(a)}_{m,+}(u) =
\epsilon^{(a)}_m(u + \log{2 \over R})$ and passed to
the limit $R \rightarrow 0$ firstly
to deduce $\epsilon^{(a)}_{m,+}(+\infty) = +\infty$
from the assumption ${\cal M}_a > 0$.
We have also set
$f^{(a)}_m =
\bigl(1 + \hbox{exp}(\epsilon^{(a)}_{m,+}(-\infty))\bigr)^{-1}$,
which is natural since the simplest branch choice
$\log f^{(a)}_m, \log(1-f^{(a)}_m) \in {\bf R}$
for all $(a,m) \in G$ in
\dter\
 then yields $D^{(a)}_m = 0$ hence the ground state value
$c = c_{\rm PF}$
by means of the dilogarithm conjecture \ccc.
However, one may allow various branches in \dter\ and
thereby introduces non-trivial chemical potentials and possibly
extracts the excitation spectra as argued in \Mae,\KLP,\KMe,\Fen.
To be more precise and systematic, we introduce
the universal covering space ${\cal R}$
of ${\bf C}\setminus \{ 0,1 \}$ and the covering map
$\ti{i}:{\cal R} \rightarrow {\bf C}\setminus \{ 0,1 \}$,
which specifies analytic continuations
of the dilogarithm.
The effective central charge $c$ \cfun\
 is then to be understood
as a function on the set of the
points $(\ti{f}\am)_{(a,m) \in G}$ on
${\cal R}^{|G|}$
such that $\ti{i}(\ti{f}\am)=f\am$, i.e.,
\eqn\gcfun{
{\pi^2 \over 6}c({\cal S}) =
\sum_{(a,m) \in G}\Bigl(
\ti{L}\bigl(\ti{f}^{(a)}_m\bigr) -
{\pi i\over 2}\ti{D}^{(a)}_m
\Log
\bigl(1-\ti{f}^{(a)}_m\bigr)\Bigr),
}
\eqn\gdter{\pi i \ti{D}^{(a)}_m=
\Log\bigl(\ti{f}^{(a)}_m\bigr) -
\sum_{(b,k) \in G}K^{m\, k}_{a\, b}
\Log
\bigl(1-\ti{f}^{(b)}_k\bigr),
}
as introduced in eq.(11) of \KN\ (with $z=0$ therein).
Here, ${\cal S}$ denotes
the collection
$({\cal C}_{a,m})_{(a,m) \in G}$
of the contours ${\cal C}_{a,m}$
from an arbitrary base point to $0 < f^{(a)}_m < 1$
in ${\bf C}\setminus \{ 0,1 \}$ specifying the point $\ti{f}\am$ on
${\cal R}$.
We warn readers that ${\cal C}_{a,m}$ here does {\it not}
mean the integration contour as opposed to the convention in
\KN.
We fix the branch $-\pi < {\rm Im}(\log(\cdot)) \le \pi$ in \dilog\
hence $L(x)$ has the cuts $(-\infty, 0]$ and $[1, +\infty)$
on the complex $x-$plane.
The $\ti{L}(\cdot)$ and $\Log(\cdot)$ in \gcfun,\gdter\
stand for the analytic continuations of $L(\cdot)$ and
$\log(\cdot)$ to ${\cal R}$, respectively.
Because $c({\cal S})$ actually depends only on the homotopy
classes of the contours ${\cal C}_{a,m}$,
we shall parametrize them by
the integers $\xi^{(a)}_{m,j},\eta^{(a)}_{m,j}\, (j \ge 1)$ as
${\cal C}_{a,m} =
{\cal C}[f^{(a)}_m \vert
\xi^{(a)}_{m,1},\xi^{(a)}_{m,2}, \ldots \vert
\eta^{(a)}_{m,1},\eta^{(a)}_{m,2}, \ldots]$, wherein
the notation
${\cal C}[f \vert
\xi_1,\xi_2, \ldots \vert
\eta_1,\eta_2, \ldots]$ signifies the contour
going from the base point to $f$ as follows (Fig.1).
It firstly
goes across the cut $[1,+\infty)$ for $\eta_1$ times then
crosses the other cut $(-\infty, 0]$
for $\xi_1$ times then
$[1,+\infty)$ again for $\eta_2$ times,
$(-\infty, 0]$ for $\xi_2$ times and so on
before approaching $f$ finally.
Here intersections have been counted as $+1$ when
the contour goes across the cut $(-\infty, 0]$ (resp. $[1,+\infty)$)
from the upper (resp. lower) half plane to the lower
(resp. upper) and $-1$ if opposite.
We call $\xi_j$ and $\eta_j$ the winding numbers and
assume that they are all
zero for $j$ sufficiently large.
{}From these definitions one deduces the
formulas
\eqn\aclog{
\Log(\ti{f}) = \log f + 2\pi i (\sum_{j \ge 1}\xi_j),\quad
\Log(1-\ti{f}) = \log (1-f) + 2\pi i (\sum_{j \ge 1}\eta_j),
}
\eqn\acdilog{
\eqalign{
\ti{L}(\ti{f}) =& L(f) +
\pi i \bigl(\sum_{j \ge 1}\xi_j \bigr) \log (1-f) -
\pi i \bigl(\sum_{j \ge 1}\eta_j \bigr) \log f \cr
& \quad
- 2\pi^2\bigl(\sum_{j \ge 1}\xi_j \bigr)
\bigl(\sum_{j \ge 1}\eta_j \bigr) + 4\pi^2 \sum_{j \ge 1}
\xi_j(\eta_1 + \cdots + \eta_j), \cr}
}
which make the dependences on the contour
${\cal C} = {\cal C}[f \vert
\xi_1,\xi_2, \ldots \vert
\eta_1,\eta_2, \ldots]$ explicit.
The collection ${\cal S}$ of the contours is
now equivalently represented by the collection of winding numbers
$(\xi^{(a)}_{m,j}, \eta^{(a)}_{m,j})_{(a,m) \in G, j \ge 1}$.
By applying \aclog, \acdilog\
 to \gcfun, \gdter\
 and
using
$\log\bigl(f^{(a)}_m\bigr) =
\sum_{(b,k) \in G}K^{m\, k}_{a\, b}
\log\bigl(1-f^{(b)}_k\bigr)$ from \feq,
one can split the $c({\cal S})$ into
${\cal S}-$dependent and independent parts.
The latter turns our to be $c_{\rm PF}$ due to \ccc\
 and we get
\footnote{$^1$}{Using this opportunity we remark that
eqs.(9b) and (12d) (with $z=0$) in \KN\
are erroneous and should be corrected
as eqs.(29) and (31) here, respectively.}
\eqn\csplit{
c({\cal S}) = c_{\rm PF} - 24T({\cal S}),
}
\eqn\tter{
T({\cal S}) = {\cal K}({\bf n}) -
\sum_{(a,m) \in G}\sum_{j \ge 1}
\xi^{(a)}_{m,j}(\eta^{(a)}_{m,1} + \cdots + \eta^{(a)}_{m,j}),
}
where ${\cal K}({\bf n})$ is defined in \propb\
 and
the vector
$\bn =( n^{(a)}_m )_{(a,m) \in G} \in {\bf Z}^{|G|}$
is specified from ${\cal S}$ by
\eqn\neta{
n^{(a)}_m = \sum_{j \ge 1} \eta^{(a)}_{m,j}.
}
The formulas \csplit-\neta\
 describe the ${\cal S}-$dependence
of the effective central charge manifestly.
\par
Let us now investigate the spectra of the $c({\cal S})$ when
${\cal S}$ consists of those contours
that intersect
the cuts $[1, +\infty)$ and $(-\infty, 0]$ always from
the lower half plane to the upper with the former crossed firstly
if ever.
Respecting \neta, such an ${\cal S}$ is a collection of
${\cal C}_{a,m}$  $((a,m) \in G)$ parametrized as
\eqn\para{
{\cal C}_{a,m} =
{\cal C}[f^{(a)}_m \vert
\xi^{(a)}_{m,1},\ldots,\xi^{(a)}_{m,n^{(a)}_m},0,0,\ldots
\vert \overbrace{1,\dots,1}^{n^{(a)}_m},0,0,\ldots]
}
for some $n^{(a)}_m \ge 0$ and
$\xi^{(a)}_{m,1},\ldots,\xi^{(a)}_{m,n^{(a)}_m} \le 0$.
Denote by ${\cal O}$ the totality of such ${\cal S}$'s.
Then from \csplit-\para, one can compute the
spectra of the effective central charge as follows.
\eqn\keisan{
\eqalign{
\sum_{{\cal S} \in {\cal O}} q^{-c(\cS)/24}= &
q^{-c_{\rm PF}/24}
\sum_{n^{(1)}_1, \dots, n^{(r)}_{\ell_r-1} \geq 0}
q^{{\cal K}({\bf n})}
\prod_{(a,m) \in G}
\sum_{\xi^{(a)}_{m,1}, \dots, \xi^{(a)}_{m,n\am} \le 0
} q^{-\sum_{j= 1}^{n\am}
j \xi^{(a)}_{m,j}}
\cr
= &
q^{-c_{{\rm PF}}/24}\sum_{\bn} {q^{{\cal K}({\bf n})}
\over (q)_{\bn}}. \cr}
}
According to \propa, the last expression is nothing but the
character ${\rm ch}(\bar\Omega^0)$.
In this way, the spectra
of the effective central charge \gcfun\
 leads to
the PF CFT character itself.
This extends our earlier observation in
\KN\ (with $\Lambda = 0$) further toward
a structural correspondence between CFTs and TBA.
Namely, the independent states in the
Hilbert space
$\bar\Omega^0$ are in one to one correspondence
to the lifts $( \ti{f}\am )_{(a,m) \in G} \in {\cal R}^{|G|}$
of the point $( f\am )_{(a,m) \in G}$ parametrized by the set $\cO$.
%
\vskip5mm \noindent
{\bf 4. Discussions}
\par\noindent
We have seen that the
whole excitation spectra in the PF vacuum module
$\Omega^0$ is obtainable from
the UV free energy (or the effective central charge $c(\cS)$)
by a certain analytic continuation procedure.
It implies that the ground state energy also possesses informations
on the excitations since the latters correspond to just
different branches of the former.
It will be interesting to seek such a phenomenon
in a wider class of models in 2d statistical mechanics and
quantum field theories.
As for our examples in this Letter,
there are at least two routes to possibly explain this
phenomenon.
The first one is to interpret $c(\cS)$ as
the expectation value of the symmetry operator under which
the corresponding excited state
becomes the leading \Fen.
Though this argument has been applied almost exclusively to some
primary excitations,
one may generalize it by including descendant
operators to accommodate the full spectra.
The second route is to regard
$c(\cS)$ as representing
finite-size corrections \BCN,\Aff\ to various transfer matrix
eigenvalues of critical RSOS-type \ABF,\DJKMONP,\JKMO\ spin chains.
In such analyses one treats the integral equations
similar to \tba\ originated from
the actual functional relations \BRIJ\ among the row to row
transfer matrices as done in \KLP\ for
$\widehat{g} = A^{(1)}_1$.
There, non trivial branch choices have indeed been
observed to yield
various eigenvalues.
Thus our prescription here might be related to
such an approach by using the $U_q(\widehat{g})$ functional
relation in \KN.
\par
The $q-$series formula \propa\ and the computation in the previous
section concern the UV limit $R \rightarrow 0$ where
the TBA just starts to deform the CFT.
So in principle they should allow a
``continuous deformation" in some sense
which will exhibit rich integrability structures away from
criticality.
\par
Finally, though we have considered only the vacuum module case
in this Letter,
it is natural to expect similar formulas to \propa\
for general PF modules in the light of the result in \KN.
We hope to report them in our
forthcoming paper \KNS.
\par
\vskip0.7cm
We would like to thank
I.\ Cherednik, P.\ Fendley,
E.\ Frenkel, S.\ Hosono and K.\ Mimachi
for valuable discussions.
T.\ N.\ would like to thank Prof.\ C.\ Vafa for his kind hospitality.
This work is supported in part by JSPS fellowship,
NSF grants PHY-87-14654, PHY-89-57162
and Packard fellowship.
\vskip0.7cm
\listrefs
\vskip2cm
\centerline{{\bf Figure Captions}}
\par
\noindent
Fig.1
\par
\noindent
An example of a contour from a base point $z_0$ to
a point  $f \in (0,1)$ in ${\bf C}\setminus \{ 0,1 \}$.
Its homotopy type is parametrized as
${\cal C}[f| -2, 0, \dots |2, 1, 0, \dots]$ or also as
${\cal C}[f| 0, -2, 0, \dots |
1, 1, 1, 0, \dots]$.
\bye